\documentclass[onecolumn,amsmath,amssymb]{revtex4}
\usepackage{soul,amsmath,amsfonts,amssymb,graphicx,natbib,color}

\def\Okpq{\varOmega_{k,pq}}
\def\Okpqp{\varOmega_{k',pq}}
\def\Okpqpp{\varOmega_{k'',pq}}
\def\Aks{A^s_k}
\def\Akss{A^{s'}_{k'}}
\def\aks{a^s_k}
\def\akss{a^{s'}_{k'}}
\def\aksss{a^{s''}_{k''}}
\def\akssss{a^{s'''}_{k'''}}
\def\Aksp{A^{s_{p}}_{p}}
\def\aksp{a^{s_{p}}_{p}}
\def\Aksq{A^{s_{q}}_{q}}
\def\aksq{a^{s_{q}}_{q}}
\def\pp{{\bm p}}
\def\qq{{\bm q}}

\def\del{\delta_{k,pq}}
\def\delp{\delta_{k',pq}}
\def\delpp{\delta_{k'',pq}}

\def\xx{{\bm x}}

\def\kk{{\bm k}}
\def\uu{{\bm u}}
\def\xx{{\bm x}}

\def\be{\begin{equation}}
\def\ee{\end{equation}}
\def\ba{\begin{eqnarray}}
\def\ea{\end{eqnarray}}
\def\ADD#1{{\textcolor{black}{#1}}}   % addition for comments!

\def \pmbtext#1{\leavevmode
     \setbox0\hbox{#1}
     \kern0,4pt \copy0 \kern-\wd0
     \kern-0,2pt \raise0,3pt \box0 }

\begin{document}

\title{{{Wave turbulence: the case of capillary waves (a review)}}}

\author{S\'ebastien Galtier}
\affiliation{Universit\'e Paris-Saclay, CNRS, Ecole polytechnique, Laboratoire de Physique des Plasmas, 
91128, Palaiseau, France}
\affiliation{Institut universitaire de France, Observatoire de Paris, SU}

\date{\today}

\begin{abstract}
Capillary waves are perhaps the simplest example to consider for an introduction to wave turbulence. Since the first paper by \citet{Zakharov1967},  
capillary wave turbulence has been the subject of many studies but a didactic derivation of the kinetic equation is still lacking. It is the objective of this 
paper to present such a derivation in absence of gravity and in the approximation of deep water. We use the Eulerian method and a Taylor expansion 
around the equilibrium elevation for the velocity potential to derive the kinetic equation. The use of directional polarities for three-wave interactions leads to 
a compact form for this equation which is fully compatible with previous work. \ADD{The exact solutions are derived with the so-called Zakharov transformation 
applied to wavenumbers and the nature of these solutions is discussed.} Experimental and numerical works done in recent decades are also reviewed. 
\end{abstract}

\maketitle

%%%%%%%%%%%%%%%%%%%%%%%%%%%%%%%%%%%%%%%%%%%%%%%%%%%%%%
\section{Introduction}
%%%%%%%%%%%%%%%%%%%%%%%%%%%%%%%%%%%%%%%%%%%%%%%%%%%%%%

Wave turbulence is about the long-time statistical behavior of a sea of weakly nonlinear waves \citep{Newell2011}.
The energy transfer between waves occurs mostly among resonant sets of waves and the resulting energy distribution, far from a 
thermodynamic equilibrium, is characterized by a wide power law spectrum and a high Reynolds number. This range of wave numbers 
is generally localized between large scales at which energy is injected in the system (sources) and small scales at which waves break 
or dissipate (sinks). 

Pioneering works on wave turbulence date back to the sixties when it was established that the stochastic initial value problem for weakly 
coupled wave systems has a natural asymptotic closure induced by the dispersive nature of the waves and the large separation of linear 
and nonlinear time scales \citep{Benney1966}. In the meantime, \citet{Zakharov1966} showed, in the case of four-wave interactions, that 
the kinetic equations derived from 
the wave turbulence analysis have exact equilibrium solutions which are the thermodynamic zero flux solutions and the finite flux solutions 
which describe the transfer of conserved quantities between sources and sinks. The solutions, first published for isotropic turbulence 
\citep{Zakharov1966,Zakharov1967} were then extended to the anisotropic case \citep{Kuznetsov1972}.  

Wave turbulence is a very common natural phenomenon found, for example, with/in 
gravity waves \citep{Falcon2007,Aubourg2016,Cazaubiel2019,Cazaubiel2019b}, 
capillary waves \citep[see also the discussion in section \ref{sec9},][]{Holt1996,Falcon2009,Berhanu2013},
quantum turbulence \citep{Lvov2003,Proment2009,Lvov2010,Proment2012}, 
nonlinear optics \citep{Dyachenko1992,Laurie2012,Picozzi2014}, 
inertial waves \citep{Galtier2003,Bellet2006,Bourouiba2008,Scott2014,Godeferd2015}, 
magnetostrophic waves \citep{Galtier2014,Menu2019}, 
elastic plates \citep{Boudaoud2008,Cobelli2009,During2006,Yokoyama2013,Chibbaro2016,Hassaini2019}, 
plasma waves \citep{Galtier2000,Galtier2006b,Galtier2006,MKG15,Passot2019}, 
or primordial gravitational waves \citep{Galtier2017,Galtier2018,GNBT2019}. 
These few examples demonstrate the vitality of the domain. In this paper we will take the example of capillary waves to present -- in a didactic 
way -- the analytical theory of weak wave turbulence. Capillary waves are perhaps the simplest example to consider for such an introduction 
to wave turbulence because they are intuitive (we can easily produce such waves) and they imply only three-wave interactions. Since the first paper 
by \citet{Zakharov1967} capillary wave turbulence has been the subject of many studies that will be discussed later (in section \ref{sec9}).
A didactic derivation of the kinetic equation is, however, still lacking. 
Therefore, it is our objective to present such a derivation in absence of gravity and in the approximation of deep water. 

\ADD{The kinetic equations of wave turbulence are obtained under some assumptions. First, we need to introduce a small parameter over which the 
development will be made. Basically, the assumption of small nonlinearities, or equivalently of weak wave amplitudes, will provide such a parameter. 
Second, the statistical spatial homogeneity will be used; it is a key element for the statistical closure. 
Third, we will assume that initially the cumulants decay sufficiently rapidly as the increments become large, which means that the fields at 
distant points are initially uncorrelated. Therefore, we do not consider initially a situation with large coherent structures. 
Fourth, the kinetic equations obtained are only valid in a bounded domain in Fourier space which corresponds to a time-scale separation 
between the wave time and the non-linear time, the former being assumed to be much smaller than the latter. 
Fifth, in the statistical development the infinite box limit will be taken formally before the small non-linear limit \citep{Nazarenko11}. }

In the next section, the fundamental equations are introduced; in particular a Taylor expansion is made around the equilibrium elevation for the 
velocity potential. Solutions of the problem are given in section \ref{sec3} with phenomenological arguments. Then, the analytical theory of weak wave 
turbulence is developed in sections \ref{sec4} and \ref{sec5}; the Eulerian method is used. The detailed energy conservation is demonstrated in 
section \ref{sec6} and the derivation of the exact solutions is exposed in section \ref{sec7} while the nature of the solution is discussed after. 
Finally, experimental and numerical results about capillary wave turbulence are exposed in section \ref{sec9} and a conclusion is proposed 
in the last section.

%%%%%%%%%%%%%%%%%%%%%%%%%%%%%%%%%%%%%%%%%%%%%%%%%%%%%%
\section{Capillary waves in equation} \label{sec2}
\begin{figure}[t]
\center
\centerline{\includegraphics[width=.8\linewidth]{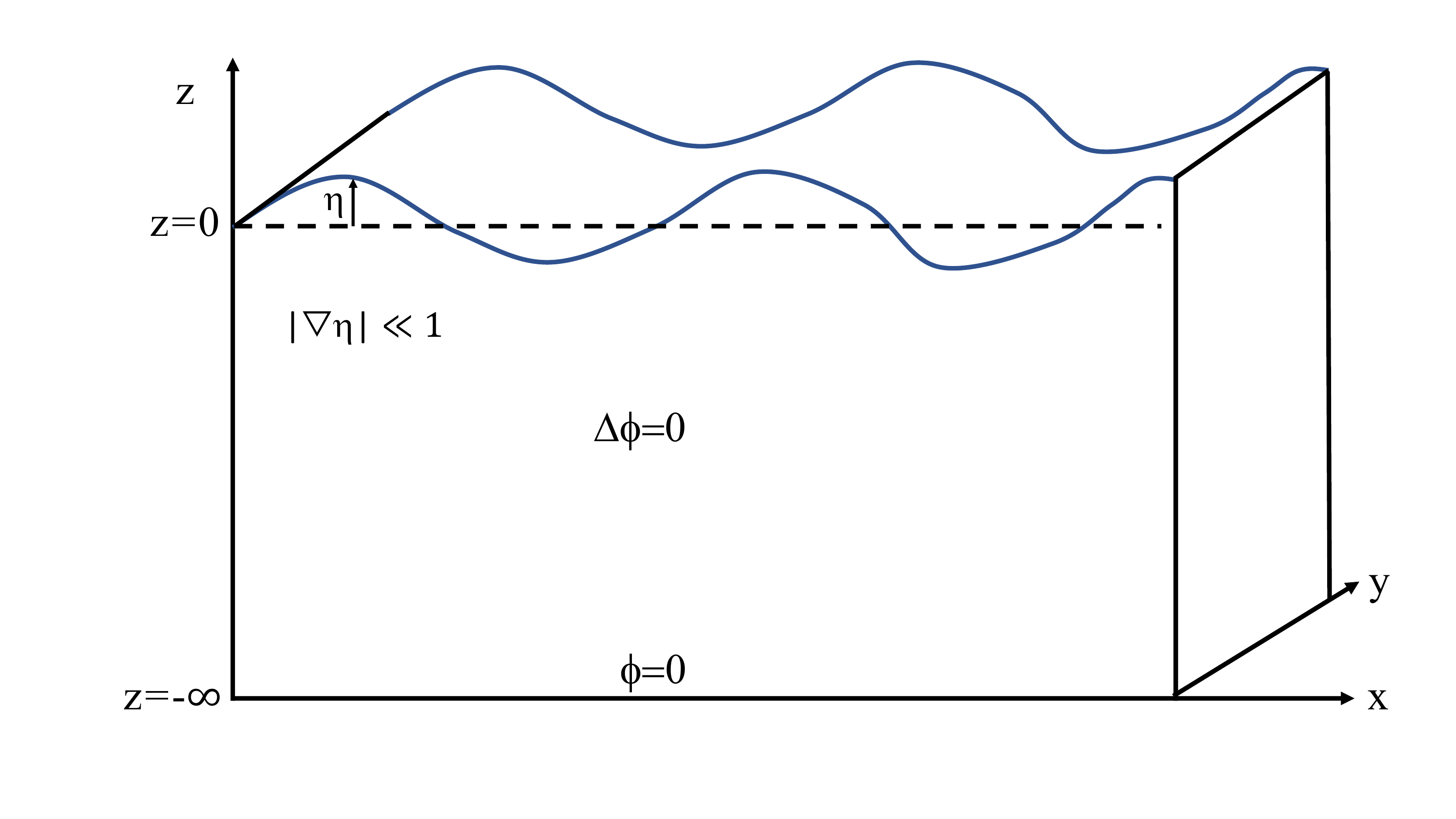}}
\caption{Schematic view of capillary waves in a deep water. It is assumed that the deformation $\eta$ of the air-water interface is on average at the 
altitude $z=0$ and is such that $\vert {\bm \nabla} \eta \vert \ll 1$, that is, of weak amplitude. In addition, we will assume that the fluid is incompressible 
(${\bm \nabla} {\bm \cdot} \uu = 0$) and irrotational ($\uu = {\bm \nabla} \phi$): in this case $\Delta \phi = 0$. The deep water hypothesis means that the potential 
$\phi$ is zero at the altitude $z=-\infty$.}
\label{Fig1}
\end{figure}

%%%%%%%%%%%%%%%%%%%%%%%%%%%%%%%%%%%%%%%%%%%%%%%%%%%%%%
With gravity waves, capillary waves constitute the most common surface waves encountered in nature. The latter have an advantage over 
the first in the sense that they are easier to treat analytically in the nonlinear regime. To introduce the physics of capillary waves we will consider 
an incompressible fluid (${\bm \nabla} {\bm \cdot} \uu = 0$) (like water) subject to irrotational movements ($\uu = {\bm \nabla} \phi$ with $\phi$ the velocity potential). 
This condition is well justified when the air-water interface is disturbed by a wind blowing unidirectionally (a typical condition encountered in the sea). 
The nonlinear equations describing the dynamics of capillary waves are obtained by first noting that the deformation of the fluid at the air-water 
interface verifies the exact Lagrangian relation:
\be
\frac{{\mathrm d} \eta}{{\mathrm d} t} = u_{z} = \frac{\partial \phi}{\partial z}\vert_{\eta} \, , 
\label{EQ1}
\ee
where $\eta(x,y,t)$ is the deformation and $\phi(x,y,z,t)$ the velocity potential (see figure \ref{Fig1} for an illustration).  
The \ADD{Bernoulli} equation (inviscid case) applied to the free surface of the liquid (at $z=\eta$) writes: 
\be
\ADD{\frac{\partial \phi}{\partial t}\vert_{\eta}  = -\frac{1}{2} \left( {\bm \nabla} \phi \right)^{2}\vert_{\eta} + \sigma \Delta \eta \, ,} \label{euler1}
\ee
where $\sigma=\gamma / \rho_{water}$ with $\gamma$ the coefficient of surface tension (for the air-water interface $\gamma \simeq 0.07$\,N/m) 
and $\rho_{water}$ the mass density of water. Note that the mass density of the air is negligible compared to that of water. 
The surface tension term is obtained by assuming that the deformation is relatively weak, i.e. $\vert {\bm \nabla} \eta \vert \ll 1$. 
This tension is responsible for a discontinuity between the fluid pressure at its free surface $P_f$ and the pressure of the atmosphere $P_a$; 
it is modeled by the relation $P_{f}-P_{a}=\sigma/R$ with $R$ the radius of curvature of the free surface \citep{Guyon2015}. 
The hypothesis of a weak deformation (or weak curvature) makes it possible to simplify the modeling.
%
%%%%%%%%%%%%%%%%%%%%%%%%%%%%%%%%%%%%%%%%%%%%%%%
%%%%%%%%%%%%%%%%%%%%%%%%%%%%%%%%%%%%%%%%%%%%%%%
%
\ADD{After developing equation (\ref{EQ1}) we obtain the system:}
\begin{subequations}
\be
\label{Z1}
\frac{\partial \eta}{\partial t} = - {\bm \nabla}_{\perp} \phi\vert_{\eta} {\bm \cdot} {\bm \nabla}_{\perp} \eta + \frac{\partial \phi}{\partial z}\vert_{\eta} \, , 
\ee
\be
\frac{\partial \phi}{\partial t}\vert_{\eta} = - \frac{1}{2} \left( {\bm \nabla} \phi \right)^{2}\vert_{\eta} + \sigma \Delta \eta \, , \label{Z2}
\ee
\end{subequations}
where the symbol $\perp$ means that we take only the space derivative in the $x$ and $y$ directions. The system (\ref{Z1})--(\ref{Z2}) is the 
one that has been used by \citet{Zakharov1967} in the limit of weak deformation to develop the theory that interests us in this paper. Basically, it 
involves the use of the potential $\phi$ at $z=\eta$ which may be difficult to manipulate. One way to overcome this difficulty is to use a Taylor 
development (at quadratic order) to express $\phi$ at $z=\eta$ from its Eulerian value at $z=0$ (equilibrium elevation); it leads to 
\ADD{\citep{Benney1962,Case1977}}:
\begin{subequations}
\be
\frac{\partial \eta}{\partial t} = - {\bm \nabla}_{\perp} \phi\vert_{0} {\bm \cdot} {\bm \nabla}_{\perp} \eta + \frac{\partial \phi}{\partial z}\vert_{0} 
+ \eta \frac{\partial^{2} \phi}{\partial z^{2}}\vert_{0} \, , \label{N1} 
\ee
\be
\frac{\partial \phi}{\partial t}\vert_{0} + \eta \frac{\partial^{2} \phi}{\partial z \partial t}\vert_{0} =
-\frac{1}{2} \left( {\bm \nabla} \phi \right)^{2}\vert_{0} + \sigma \Delta \eta \, . \label{N2}
\ee
\end{subequations}
We have limited ourselves to quadratic nonlinearities because the problem of capillary wave turbulence can be solved at this level, i.e. for three-wave 
interactions \ADD{\citep{McGoldrick1965}}. This situation differs from the case of gravity waves which must be treated at the cubic level and so with four-wave 
interactions. Equations (\ref{N1})--(\ref{N2}) are those we will consider to develop the theory of capillary wave turbulence \ADD{\citep{Newell2013}}. 
They are complemented by the incompressible and irrotational conditions of the fluid:
\be
\Delta \phi = 0 \, .
\ee
Under the deep water hypothesis, i.e $\phi=0$ at $z=-\infty$ (see figure \ref{Fig1}), we get a function of the form:
\be
\phi(x,y,z,t) = \psi(x,y,t) {\mathrm e}^{kz} \, , 
\ee
with $k$ the norm of the wave vector $\kk \equiv (k_x,k_y)$. 

The linearization of equations (\ref{N1})--(\ref{N2}) will give us the dispersion relation. We obtain after a Fourier transform:
\begin{subequations}
\be
-{\mathrm i} \omega_{k} \hat \eta_{k} = k \hat \phi_{k} \, , 
\ee
\be
-{\mathrm i} \omega_{k} \hat \phi_{k} = - \sigma k^{2} \hat \eta_{k} \, ,
\ee
\end{subequations}
with by definition:
\begin{subequations}
\be
\hat \eta_{k} \equiv \hat \eta(k_{x},k_{y}) = \frac{1}{(2\pi)^2} \int \eta(\xx) {\mathrm e}^{-i \kk {\bm \cdot} \xx} {\mathrm d}\xx\, , 
\ee
\be
\hat \phi_{k} \equiv \hat \phi(k_{x},k_{y}) = \frac{1}{(2\pi)^2} \int \phi(\xx) {\mathrm e}^{-i \kk {\bm \cdot} \xx} {\mathrm d}\xx\, . 
\ee
\end{subequations}
Finally, we obtain the dispersion relation:
\be
\omega^{2}_{k} = \sigma k^{3} \, .
\ee
Note that the presence of gravity at the linear level brings a correction to this relation with $\omega^{2}_{k} = \sigma k^{3} + gk$. 
Therefore, our study is valid in the case where $k \ADD{\gg} k_{*}$ with $k_{*} \equiv \sqrt{g \rho_{water} /\gamma}$ (we have explicitly written 
the coefficient of surface tension to obtain a numerical value). This corresponds to a critical wavelength $\lambda_{*} \simeq 1.7$\,cm 
for the air-water interface. As a result, capillary waves appear at small scales. They are dispersive with a phase velocity $v_{\phi}$ which 
increases with the wave number ($v_{\phi} \propto \sqrt{k}$). This property can be observed by disturbing the surface of the water: the 
waves of small wavelengths are the fastest to escape from the disturbed region. Note in passing that in the case of gravity waves 
we have the inverse situation (easily verified by the experiment): it is the gravity waves of long wavelengths which escape the fastest 
of the disturbed region (but they are preceded by capillary waves).

%%%%%%%%%%%%%%%%%%
\section{Phenomenology}\label{sec3}
%%%%%%%%%%%%%%%%%%

Phenomenology plays a fundamental role in turbulence because in the regime of strong turbulence it is the method used to reach a spectral 
prediction, for example, for the spectrum of energy. In the case of wave turbulence, it is possible to obtain analytically this solution, however, 
the phenomenological analysis remains indispensable, on the one hand, to arrive quickly to a first prediction and, on the other hand, to be 
able to explain simply how emerges the solution that we are looking for. If we take the studied equation (\ref{N2}) and we only retain the 
nonlinear contribution, we arrive at the following phenomenological expression:
\be
\frac{\phi}{\tau_{NL}} \sim k^2 \phi^2 \, ,  \label{pheno1}
\ee
where $\tau_{NL}$ is the nonlinear time with ${\partial \phi}/{\partial t} \sim \phi/ \tau_{NL}$ and $\left( {\bm \nabla} \phi \right)^{2} \sim k^{2} \phi^2$. 
Then, we obtain:
\be
\tau_{NL} \sim \frac{1}{k^2 \phi} \, .  \label{pheno2}
\ee
A similar analysis done from equation (\ref{N1}) gives the same expression. It can be noted, however, that the nonlinear term 
$\eta {\partial^{2} \phi}/({\partial t \partial z})\vert_{0}$ of equation (\ref{N2}) gives an additional expression when it is balanced with the temporal 
derivative (one must also use expression (\ref{pheno2})), namely:
\be
k\phi^2 \sim k^2 \eta^2 \, .
\ee
This can be interpreted as an equipartition relation between the kinetic ($k\phi^2$) and potential ($k^2 \eta^2$) energies. To find a prediction 
for the total energy spectrum, we have to introduce the mean energy transfer rate \ADD{(in wavenumbers)} 
$\varepsilon$ in the inertial range as follows:
\be
\varepsilon \sim \frac{kE_{k}}{\tau_{tr}} \sim \frac{kE_{k}}{\omega \tau_{NL}^2} \sim \frac{kE_{k}k^{4} \phi^{2}}{k^{3/2}} \sim k^{7/2} E_k^2 \, ,
\label{relateps}
\ee
with $\tau_{tr}$ the transfer (or cascade) time for capillary wave turbulence and $E_{k}$ the one-dimensional total energy spectrum 
(we shall assume that this turbulence is statistically isotropic). 
\ADD{We use here the expression of $\tau_{tr}$ for triadic interactions whose expression is non-trivial to guess. Basically, the phenomenology 
behind this expression is the collision between wave-packets: the multiplicity of collisions leads to the deformation of wave-packets and eventually 
to the transfer of energy towards smaller scales. Let $\tau_{c} \sim \ell / (\omega/k) \sim 1/ \omega$ be the characteristic duration of a single collision 
between two wave packets of length $\ell$. Then, the small deformation after a single collision can be estimated through the evolution of the 
velocity potential:
\be
\phi_{\ell} (t+\tau_{c}) \sim \phi_{\ell} (t) + \tau_{c} \frac{\partial \phi_{\ell}}{\partial t} \sim \phi_{\ell} (t) + \tau_{c} \frac{\phi_{\ell}^{2}}{\ell^{2}} \, .
\ee
In this expression, the nonlinear term is evaluated from equation (\ref{N2}). Therefore, the deformation of a wave-packet after one collision is:
\be
\Delta_{1} \phi_{\ell} \sim \tau_{c} \frac{\phi_{\ell}^{2}}{\ell^{2}} \, . 
\ee
This deformation will grow with time and for $N$ stochastic collisions the cumulative effect can be evaluated in the same manner as a 
random walk:
\be
\sum_{i=1}^{N} \Delta_{i} \phi_{\ell} \sim \tau_{c} \frac{\phi_{\ell}^{2}}{\ell^{2}}  \sqrt{\frac{t}{\tau_{c}}} \, .
\ee
The transfer time $\tau_{tr}$ corresponds to a cumulative deformation of order one, i.e. of the order of the wave-packet itself. Then:
\be
\phi_{\ell} \sim \tau_{c} \frac{\phi_{\ell}^{2}}{\ell^{2}}  \sqrt{\frac{\tau_{tr}}{\tau_{c}}} \, ,
\ee
from which we obtain:
\be
\tau_{tr} \sim \frac{1}{\tau_{c}} \frac{\ell^{4}}{\phi_{\ell}^{2}} \sim \frac{\tau^{2}_{NL}}{\tau_{c}} \sim \omega \tau^{2}_{NL} \, . 
\ee
}
From expression (\ref{relateps}), we eventually find the relationship:
\be
E_{k} \sim \sqrt{\varepsilon} k^{-7/4} \, . \label{KZsol}
\ee
From this prediction and the information about the equipartition between the kinetic and potential energies, we obtain the spectra:
\be
E_k^{\phi} \equiv \vert \phi_{k} \vert^{2} \sim \sqrt{\varepsilon} k^{-11/4} 
\quad \rm{and} \quad 
E_k^{\eta} \equiv \vert \eta_{k} \vert^{2} \sim \sqrt{\varepsilon} k^{-15/4} \, .  \label{KZsol2}
\ee
These spectra can also be written as a function of the frequency using the dispersion equation $\omega \sim k^{3/2}$. 
With the dimensional relation $k E_{k} \sim \omega E_{\omega}$, we obtain for the total energy:
\be
E_{\omega} \sim \sqrt{\varepsilon} \omega^{-3/2} \, , \label{KZsol3}
\ee
and then:
\be
E_{\omega}^{\phi} \equiv \vert \phi_{\omega} \vert^{2} \sim \sqrt{\varepsilon} \omega^{-13/6} 
\quad \rm{and} \quad 
E_{\omega}^{\eta} \equiv \vert \eta_{\omega} \vert^{2} \sim \sqrt{\varepsilon} \omega^{-17/6} \, . \label{KZsol4}
\ee
The last expression is often used for the comparison with the experiment (or the direct numerical simulation) because it is easily accessible. 

We will see later that the energy spectrum (\ref{KZsol}) can be obtained analytically as an exact solution of the capillary wave turbulence equations. 
The analytical approach also makes it possible to demonstrate that the energy cascade is forward (with a positive flux) and to estimate the so-called 
Kolmogorov constant allowing to substitute the sign '$\sim$' into '$=$' in expression (\ref{KZsol}). 

A last comment can be done about the regime of weak capillary wave turbulence if we write the ratio $\chi$ between the wave period and the nonlinear time.
With the prediction (\ref{KZsol2}), we get:
\be
\chi = \frac{1/\omega}{\tau_{NL}} \sim \frac{k^2 \phi}{\omega} \sim k^{-3/8} \, , 
\ee
which means that this turbulence becomes weaker at smaller scales. In other words, if at a given scale turbulence is weak it remains weak with 
a direct cascade. Note that the situation can be different in other systems: for example, in magnetohydrodynamics the ratio $\chi$ increases with the 
wavenumber and the inertial range of weak turbulence is therefore limited (dissipation at small scale brings of course another limit) \citep{MGK16}.
\ADD{It is also the case for gravity waves where $\chi$ increases towards small scales \citep{Nazarenko11}.}

%%%%%%%%%%%%%%%%%%
\section{Analytical theory: fundamental equation}\label{sec4}
%%%%%%%%%%%%%%%%%%

For the nonlinear treatment of capillary wave turbulence, we will move to the Fourier space and use extensively the properties of the Fourier transform. 
The system (\ref{N1})--(\ref{N2}) becomes:
\begin{subequations}
\be
\frac{\partial \hat \eta_{k}}{\partial t} -k \hat \phi_{k} =
\int [(\pp {\bm \cdot} \qq) \hat \phi_p \hat \eta_q + p^2 \hat \phi_{p} \hat \eta_{q} ]\delta(\kk-\pp-\qq) {\mathrm d}\pp {\mathrm d}\qq \, , \label{NF1} 
\ee
\be
\frac{\partial \hat \phi_{k}}{\partial t} + \sigma k^2 \hat \eta_{k} = \frac{1}{2} \int [(\pp {\bm \cdot} \qq - pq) \hat \phi_p \hat \phi_q 
+ 2 \sigma p^{3} \hat \eta_{p} \hat \eta_{q} ] \delta(\kk-\pp-\qq) {\mathrm d}\pp {\mathrm d}\qq \, . \quad \quad \quad \label{NF2}
\ee
\end{subequations}
The convolution product is expressed through the presence of the Dirac $\delta(\kk-\pp-\qq)$. 
\ADD{Note a difference with the equations obtained by \cite{Zakharov1967} whose origin can be attributed to the Taylor expansion made here around 
the equilibrium elevation.}
We now introduce the canonical variables $\Aks$ of this system:
\begin{subequations}
\be
\hat \eta_{k} \equiv \left(\frac{4}{\sigma k}\right)^{1/4} \sum_{s} \Aks \, , \label{can1}
\ee
\be
\hat \phi_{k} \equiv -{\mathrm i} (4 \sigma k)^{1/4} \sum_{s} s \Aks \, , \label{can2}
\ee
\end{subequations}
with the directional polarity $s=\pm$. Since the functions $\eta$ are $\phi$ real, we have:
\be
\hat \eta_{-k}^{*} = \hat \eta_{k} \, , \quad \hat \phi_{-k}^{*} = \hat \phi_{k} \, , 
\ee
which gives the remarkable relation: ${\Aks}^*=A_{-k}^{-s}$ (with $^{*}$ the complexe conjugate).
The introduction of expressions (\ref{can1})--(\ref{can2}) into (\ref{NF1})--(\ref{NF2}) gives (we also use the triadic relation $\qq=\kk-\pp$ to simplify 
the nonlinear term):
\ba
&&\frac{\partial \Aks}{\partial t} + {\mathrm i}s \omega_{k} \Aks = \frac{1}{2}\left(\frac{\sigma k}{4}\right)^{1/4} 
\int (\kk {\bm \cdot} \pp) \hat \phi_p \hat \eta_q \delta(\kk-\pp-\qq) {\mathrm d}\pp {\mathrm d}\qq \nonumber \\
&+& \frac{{\mathrm i}s}{4} \left(\frac{1}{4 \sigma k}\right)^{1/4} \int [(\pp {\bm \cdot} \qq - pq) \hat \phi_p \hat \phi_q 
+ 2 \sigma p^{3} \hat \eta_{p} \hat \eta_{q} ] \delta(\kk-\pp-\qq) {\mathrm d}\pp {\mathrm d}\qq \, , 
\ea
with $\omega_{k} = \sqrt{\sigma k^{3}}$.
We immediately see the relevance of the choice of definitions of canonical variables at the linear level: the left-hand side term makes explicit the dispersion 
relation. The introduction of these variables at the nonlinear level gives us:
\ba
&&\frac{\partial \Aks}{\partial t} + {\mathrm i}s \omega_{k} \Aks = \frac{-{\mathrm i} \sigma^{1/4}}{\sqrt{2}}
\int \sum_{s_{p} s_{q}} s_{p} (\kk {\bm \cdot} \pp) \left(\frac{pk}{q}\right)^{1/4} \Aksp \Aksq \delta(\kk-\pp-\qq) {\mathrm d}\pp {\mathrm d}\qq \nonumber \\
&&- \frac{{\mathrm i} s \sigma^{1/4}}{2 \sqrt{2}} \int \sum_{s_{p} s_{q}} \left[ s_p s_q (\pp {\bm \cdot} \qq - pq) \left(\frac{pq}{k}\right)^{1/4}
- \frac{2 p^3}{(kpq)^{1/4}} \right] \Aksp \Aksq \delta(\kk-\pp-\qq) {\mathrm d}\pp {\mathrm d}\qq \, . \quad  
\ea
This expression can not be used as such for the statistical development: it is necessary to simplify it and especially to make it as symmetrical 
as possible in order to facilitate the subsequent work \ADD{(it is always easier to manipulate or simplify symmetric equations)}.
The first remark concerns the first and last nonlinear terms that can be symmetrized by interchanging the wave vectors $\pp$ and $\qq$, and 
the associated polarizations $s_{p}$ and $s_{q}$; it gives us:
\ba
&&\frac{\partial \Aks}{\partial t} + {\mathrm i}s \omega_{k} \Aks = -\frac{{\mathrm i} \sigma^{1/4}}{2 \sqrt{2}}
\int \sum_{s_{p} s_{q}}  s_{p} s_{q} \Aksp \Aksq \delta(\kk-\pp-\qq) \nonumber \\
&& \left[ s (\pp {\bm \cdot} \qq - pq) \left(\frac{pq}{k}\right)^{1/4} + 
s_{q} (\kk {\bm \cdot} \pp) \left(\frac{pk}{q}\right)^{1/4} + s_{p} (\kk {\bm \cdot} \qq) \left(\frac{qk}{p}\right)^{1/4} 
- \frac{ss_{p}s_{q} (p^3+q^3)}{(kpq)^{1/4}} \right] {\mathrm d}\pp {\mathrm d}\qq . \quad  
\ea
Then we introduce and subtract several terms and get:
\ba
&&\frac{\partial \Aks}{\partial t} + {\mathrm i}s \omega_{k} \Aks = - \frac{{\mathrm i} \sigma^{1/4}}{2 \sqrt{2}}
\int \sum_{s_{p} s_{q}}  s_{p} s_{q} \Aksp \Aksq \delta(\kk-\pp-\qq) \nonumber \\
&&\left[ s (\pp {\bm \cdot} \qq + pq) \left(\frac{pq}{k}\right)^{1/4} + 
s_{q} (\kk {\bm \cdot} \pp - kp) \left(\frac{pk}{q}\right)^{1/4} + s_{p} (\kk {\bm \cdot} \qq - kq) \left(\frac{qk}{p}\right)^{1/4} \right.  \nonumber \\
&&- \left. 2spq \left(\frac{pq}{k}\right)^{1/4} + s_{q} kp \left(\frac{pk}{q}\right)^{1/4} + s_{p} kq \left(\frac{qk}{p}\right)^{1/4} 
- \frac{s s_{p}s_{q} (p^3+q^3)}{(kpq)^{1/4}} \right] {\mathrm d}\pp {\mathrm d}\qq \, .  \label{NFA}
\ea
We will see that the terms of the last line do not contribute to the nonlinear dynamics over the long times. 
For that, we introduce the frequency $\omega_{k}=\sqrt{\sigma k^{3}}$; we can then show that:
\ba
&&s_{p}s_{q} \left[-2spq \left(\frac{pq}{k}\right)^{1/4} + s_{q} kp \left(\frac{pk}{q}\right)^{1/4} + s_{p} kq \left(\frac{qk}{p}\right)^{1/4} 
- \frac{s s_{p}s_{q} (p^3+q^3)}{(kpq)^{1/4}} \right] =  \nonumber \\
&& \frac{s (s_{p}\omega_{p}+s_{q} \omega_{q})(s\omega_{k}-s_{p}\omega_{p}-s_{q}\omega_{q})}{\sigma (kpq)^{1/4}} \, .
\ea
Finally, we obtain:
\ba
&&\frac{\partial \Aks}{\partial t} + {\mathrm i}s \omega_{k} \Aks = - \frac{{\mathrm i} \sigma^{1/4}}{2 \sqrt{2}}
\int \sum_{s_{p} s_{q}}  s_{p} s_{q} \Aksp \Aksq \delta(\kk-\pp-\qq) \nonumber \\
&&\left[ s (\pp {\bm \cdot} \qq + pq) \left(\frac{pq}{k}\right)^{1/4} + s_{q} (\kk {\bm \cdot} \pp - kp) \left(\frac{pk}{q}\right)^{1/4} 
+ s_{p} (\kk {\bm \cdot} \qq - kq) \left(\frac{qk}{p}\right)^{1/4} \right] {\mathrm d}\pp {\mathrm d}\qq \nonumber \\
&&-\frac{{\mathrm i} s\sigma^{1/4}}{2 \sqrt{2}} \int \sum_{s_{p} s_{q}} 
 \frac{(s_{p}\omega_{p}+s_{q} \omega_{q})(s\omega_{k}-s_{p}\omega_{p}-s_{q}\omega_{q})}{\sigma (kpq)^{1/4}} 
\Aksp \Aksq \delta(\kk-\pp-\qq) {\mathrm d}\pp {\mathrm d}\qq \, .  
\ea
Since the amplitude of the waves is supposed to be weak, the linear terms will first of all dominate the dynamics with a variation of the phase only. 
At large times, the nonlinear terms will no longer be negligible and will modify the amplitude of the waves. Under these conditions, it is relevant for 
the canonical variables to separate the amplitude from the phase. We introduce a small parameter $\epsilon \ll 1$ and write:
\be
\Aks \equiv \epsilon \aks {\mathrm e}^{-{\mathrm i} s \omega_{k}t} \, ,
\ee
hence the expression:
\ba
&&\frac{\partial \aks}{\partial t} = - \frac{{\mathrm i} \epsilon \sigma^{1/4}}{2 \sqrt{2}} 
\int \sum_{s_{p} s_{q}}  s_{p} s_{q} \aksp \aksq \delta(\kk-\pp-\qq) {\mathrm e}^{{\mathrm i}(s \omega_{k} - s_{p}\omega_{p} -s_{q}\omega_{q})t} \nonumber \\
&&\left[ s (\pp {\bm \cdot} \qq + pq) \left(\frac{pq}{k}\right)^{1/4} + s_{q} (\kk {\bm \cdot} \pp - kp) \left(\frac{pk}{q}\right)^{1/4} 
+ s_{p} (\kk {\bm \cdot} \qq - kq) \left(\frac{qk}{p}\right)^{1/4} \right] {\mathrm d}\pp {\mathrm d}\qq \nonumber \\
&&-\frac{{\mathrm i} \epsilon s\sigma^{1/4}}{2 \sqrt{2}} \int \sum_{s_{p} s_{q}} 
 \frac{(s_{p}\omega_{p}+s_{q} \omega_{q})(s\omega_{k}-s_{p}\omega_{p}-s_{q}\omega_{q})}{\sigma (kpq)^{1/4}} 
\aksp \aksq {\mathrm e}^{{\mathrm i}(s \omega_{k} - s_{p}\omega_{p} -s_{q}\omega_{q})t} \delta(\kk-\pp-\qq) {\mathrm d}\pp {\mathrm d}\qq \, .  \quad \quad 
\label{NFFs} 
\ea
We will be interested in the nonlinear dynamics that emerges at long time. By long time, we mean a time $\tau$ much longer than the wave period, 
that is $\tau \gg 1/\omega_{k}$. It is clear that the relevant contributions are those that cancel the coefficient in the exponential. As a result, the 
secular contributions will not be provided by the second integer of equation (\ref{NFFs}) which exactly cancels for this condition. We will therefore 
neglect this term afterwards. Note, however, that this term does not appear in the derivation made by \citet{Zakharov1967}. 
Finally, we obtain the following nonlinear equation for the evolution of capillary wave amplitude:
\ba
&&\frac{\partial \aks}{\partial t} = {\mathrm i} \epsilon \int \sum_{s_{p} s_{q}}  L_{-kpq}^{-ss_{p}s_{q}} 
\aksp \aksq {\mathrm e}^{{\mathrm i}\Okpq t} \del {\mathrm d}\pp {\mathrm d}\qq  \, , \quad \quad \label{fonda1}
\ea
with by definition $\Okpq \equiv s \omega_{k} - s_{p}\omega_{p} -s_{q}\omega_{q}$, $\del \equiv \delta(\kk-\pp-\qq)$ and
\ba
&&L_{kpq}^{ss_{p}s_{q}}  \equiv \nonumber \\
&&\frac{s_{p} s_{q} \sigma^{1/4}}{2 \sqrt{2}} \left[s(\pp {\bm \cdot} \qq + pq) \left(\frac{pq}{k}\right)^{1/4} 
+ s_{p} (\kk {\bm \cdot} \qq + kq) \left(\frac{qk}{p}\right)^{1/4} + s_{q} (\kk {\bm \cdot} \pp + kp) \left(\frac{pk}{q}\right)^{1/4} \right] \, . 
\ea
Equation (\ref{fonda1}) governs the slow evolution of capillary waves of weak amplitude. It is a quadratic nonlinear equation: these nonlinearities 
correspond to the interactions between waves propagating in the directions $\pp$ and $\qq$, and in the positive ($s_{p}$, $s_{q}>0$) 
or negative ($s_{p}, s_{q} <0$) direction. 
Equation (\ref{fonda1}) is fundamental for our problem since it is from this that we will make a statistical development on asymptotically long times. 
This development is based on the symmetries of the fundamental equation: a lack of symmetry can be a source of failure \ADD{in the sense that 
the development is heavy and the chance to make a mistake increases seriously if the equations are less symmetric which means in general bigger.
Furthermore, several simplifications appear clearly when equations are symmetric.}
In our case, the interaction coefficient verifies the following symmetries:
\begin{subequations}
\be
L_{kpq}^{ss_{p}s_{q}} = L_{kqp}^{ss_{q}s_{p}} \, , 
\ee
\be
L_{0pq}^{ss_{p}s_{q}} = 0 \, , 
\ee
\be
L_{-k-p-q}^{ss_{p}s_{q}} = L_{kpq}^{ss_{p}s_{q}} \, , 
\ee
\be
L_{kpq}^{-s-s_{p}-s_{q}} = - L_{kpq}^{ss_{p}s_{q}} \, , 
\ee
\be
ss_{q} L_{qpk}^{s_{q}s_{p}s} = L_{kpq}^{ss_{p}s_{q}} \, , 
\ee
\be
ss_{p} L_{pkq}^{s_{p}ss_{q}} = L_{kpq}^{ss_{p}s_{q}} \, .
\ee
\end{subequations}
These symmetries are sufficient in number for the success of the statistical study.

%%%%%%%%%%%%%%%%%%
\section{Analytical theory: statistical approach}\label{sec5}
%%%%%%%%%%%%%%%%%%

We now move on to a statistical description. We use the ensemble average $\langle ... \rangle$ and we define the following spectral correlators 
(cumulants) for homogeneous turbulence:
\begin{subequations}
\be
\langle \aks \akss \rangle = q^{ss'}_{kk'}(\kk,\kk') \delta(\kk+\kk') \, , \label{stat1} 
\ee
\be
\langle \aks \akss \aksss \rangle = q^{ss's''}_{kk'k''}(\kk,\kk',\kk'') \delta(\kk+\kk'+\kk'') \, , \label{stat2}  
\ee
\be
\langle \aks \akss \aksss \akssss \rangle = q^{ss's''s'''}_{kk'k''k'''}(\kk,\kk',\kk'',\kk''') \delta(\kk+\kk'+\kk''+\kk''') \nonumber 
\ee
\be
+ q^{ss'}_{kk'}(\kk,\kk') q^{s''s'''}_{k''k'''}(\kk'',\kk''') \delta(\kk+\kk') \delta(\kk''+\kk''') \nonumber
\ee
\be
+ q^{ss''}_{kk''}(\kk,\kk'') q^{s's'''}_{k'k'''}(\kk',\kk''') \delta(\kk+\kk'') \delta(\kk'+\kk''') \nonumber
\ee
\be
+ q^{ss'''}_{kk'''}(\kk,\kk''') q^{s's''}_{k'k''}(\kk',\kk'') \delta(\kk+\kk''') \delta(\kk'+\kk'')  \, . \label{stat3}
\ee
\end{subequations}
From the fundamental equation (\ref{fonda1}), we obtain:
\ba
\frac{\partial \langle \aks \akss \rangle}{\partial t} &=& 
\left\langle \frac{\partial \aks}{\partial t} \akss \right\rangle + \left\langle \aks \frac{\partial \akss}{\partial t} \right\rangle \nonumber \\
&=& {\mathrm i} \epsilon \int \sum_{s_{p} s_{q}}  L_{-kpq}^{-ss_{p}s_{q}} \langle \akss \aksp \aksq \rangle 
{\mathrm e}^{{\mathrm i}\Okpq t} \del {\mathrm d}\pp {\mathrm d}\qq \nonumber \\
&+& {\mathrm i} \epsilon \int \sum_{s_{p} s_{q}}  L_{-k'pq}^{-s's_{p}s_{q}} \langle \aks \aksp \aksq \rangle 
{\mathrm e}^{{\mathrm i}\Okpqp t} \delp {\mathrm d}\pp {\mathrm d}\qq  \, . \label{2ndo}
\ea
At next order, we have:
\ba
\frac{\partial \langle \aks \akss \aksss \rangle}{\partial t} &=& 
\left\langle \frac{\partial \aks}{\partial t} \akss \aksss \right\rangle 
+ \left\langle \aks \frac{\partial \akss}{\partial t} \aksss \right\rangle 
+ \left\langle \aks \akss \frac{\partial \aksss}{\partial t} \right\rangle \nonumber \\
&=& {\mathrm i} \epsilon \int \sum_{s_{p} s_{q}}  L_{-kpq}^{-ss_{p}s_{q}} \langle \akss \aksss \aksp \aksq \rangle 
{\mathrm e}^{{\mathrm i} \Okpq t} \del {\mathrm d}\pp {\mathrm d}\qq \nonumber \\
&+& {\mathrm i}\epsilon \int \sum_{s_{p} s_{q}}  L_{-k'pq}^{-s's_{p}s_{q}} \langle \aks \aksss \aksp \aksq \rangle 
{\mathrm e}^{{\mathrm i} \Okpqp t} \delp {\mathrm d}\pp {\mathrm d}\qq \nonumber \\
&+& {\mathrm i} \epsilon \int \sum_{s_{p} s_{q}}  L_{-k''pq}^{-s''s_{p}s_{q}} \langle \aks \akss \aksp \aksq \rangle 
{\mathrm e}^{{\mathrm i} \Okpqpp t} \delpp {\mathrm d}\pp {\mathrm d}\qq  \, . \label{cumul3}
\ea
We are dealing here with the classical problem of closure: a hierarchy of statistical equations of increasing order is emerging. Unlike the strong 
turbulence regime, in the weak wave turbulence regime we can use the scale separation in time to achieve a natural closure of the system
\citep{Benney1966}. We insert expressions (\ref{stat1})--(\ref{stat3}) into equation (\ref{cumul3}):
\ba
&&\frac{\partial q^{ss's''}_{kk'k''}(\kk,\kk',\kk'')}{\partial t} \delta(\kk+\kk'+\kk'') = \nonumber \\
&& {\mathrm i} \epsilon \int \sum_{s_{p} s_{q}}  L_{-kpq}^{-ss_{p}s_{q}} 
[q^{s's''s_{s_{p}}s_{s_{q}}}_{k'k''pq}(\kk',\kk'',\pp,\qq) \delta(\kk'+\kk''+\pp+\qq) \nonumber \\
&&+ q^{s's''}_{k'k''}(\kk',\kk'') q^{s_{p}s_{q}}_{pq}(\pp,\qq) \delta(\kk'+\kk'') \delta(\pp+\qq) \nonumber \\
&& + q^{s's_{p}}_{k'p}(\kk',\pp) q^{s''s_{q}}_{k''q}(\kk'',\qq) \delta(\kk'+\pp) \delta(\kk''+\qq) \nonumber \\
&&+ q^{s's_{q}}_{k'q}(\kk',\qq) q^{s''s_{p}}_{k''p}(\kk'',\pp) \delta(\kk'+\qq) \delta(\kk''+\pp) ] {\mathrm e}^{{\mathrm i} \Okpq t} \del {\mathrm d}\pp {\mathrm d}\qq \nonumber \\
&&+ \, {\mathrm i}\epsilon \int \Big\{ (\kk,s) \leftrightarrow (\kk',s') \Big\} {\mathrm d}\pp {\mathrm d}\qq \nonumber \\
&&+ \, {\mathrm i} \epsilon \int \Big\{ (\kk,s) \leftrightarrow (\kk'',s'') \Big\} {\mathrm d}\pp {\mathrm d}\qq \, ,  \label{q4t1} 
\ea
where the last two lines correspond to the exchange at the level of the notations between $\kk$, $s$ in the developed expression and 
$\kk'$, $s'$ (before last line), then $\kk''$, $s''$ (last line).

We will now integrate expression (\ref{q4t1}) over both $\pp$ and $\qq$, and over the time (we assume the absence of coherent structures initially) 
considering a time of integration long compared to the 
time of reference, i.e. the period of the capillary wave\footnote{Note that it is at this level of analysis that the closures used in turbulence are different. 
Researches conducted mainly in the years 60--70 \citep{Orszag1970} led in particular to a popular closure called EDQNM (Eddy Damped Quasi-Normal 
Markovian) which makes a link in a {\it ad hoc} way between the fourth-order cumulant and the third-order moment: the fourth-order cumulant plays 
the role of damping for the third-order moment without memory effect.}. 
The presence of several Dirac functions makes it possible to conclude that the second term in the right-hand side (in the main expression) gives no 
contribution since it corresponds to $k = 0$ for which the coefficient of interaction is null; it is a property of statistical homogeneity. 
The last two terms in the right-hand side (still in the main expression) have a strong constraint on the wave vectors $\pp$ and $\qq$ which must be equal to
$-\kk'$ and $-\kk''$, respectively. 
For the fourth-order cumulant, the constraint is much less strong since only the sum of $\pp$ and $\qq$ is imposed. The implication is that for long 
times this term will not contribute to the nonlinear dynamics \citep{Benney1966}.
Finally, for long times, the second-order cumulants are only relevant when the associated polarities have different signs. To understand this, we 
must go back to the definition of the moment:
\be 
\langle \Aks \Akss \rangle = \epsilon^{2} \langle \aks \akss \rangle {\mathrm e}^{-{\mathrm i}(s\omega_{k}+s'\omega_{k'})t} \, ,
\ee
from which we see that a non-zero contribution is possible for homogeneous turbulence ($\kk=-\kk'$) only if $s=-s'$ (thus the coefficient of the 
exponential vanishes). We finally get:
\ba
&&q^{ss's''}_{kk'k''}(\kk,\kk',\kk'') \delta(\kk+\kk'+\kk'') ={\mathrm i} \epsilon \Delta(\varOmega_{kk'k''}) \delta(\kk+\kk'+\kk'') \nonumber \\
&& \left\{ \left[ L_{-k-k'-k''}^{-s-s'-s''} + L_{-k-k''-k'}^{-s-s''-s'} \right]  q^{s''-s''}_{k''-k''}(\kk'',-\kk'') q^{s'-s'}_{k'-k'}(\kk',-\kk') \right. \nonumber \\
&& +\left[ L_{-k'-k-k''}^{-s'-s-s''} + L_{-k'-k''-k}^{-s'-s''-s} \right]  q^{s''-s''}_{k''-k''}(\kk'',-\kk'') q^{s-s}_{k-k}(\kk,-\kk) \nonumber \\
&& \left. + \left[ L_{-k''-k'-k}^{-s''-s'-s} + L_{-k''-k-k'}^{-s''-s-s'} \right]  q^{s-s}_{k-k}(\kk,-\kk) q^{s'-s'}_{k'-k'}(\kk',-\kk') \right\} \, , \label{q4t2}
\ea
with:
\be
\Delta(\varOmega_{kk'k''}) = \int_0^{t\gg1/\omega} {\mathrm e}^{{\mathrm i} \varOmega_{kk'k''}t^\prime} {\mathrm d}t^\prime 
=  {{\mathrm e}^{{\mathrm i} \varOmega_{kk'k''}t} - 1 \over {\mathrm i} \varOmega_{kk'k''} } \, .
\ee
We can now write unambiguously: $q^{s-s}_{k-k}(\kk,-\kk) = q^s_k(\kk)$. Using the symmetry relations of the interaction coefficient, we obtain:
\ba
&&q^{ss's''}_{kk'k''}(\kk,\kk',\kk'') \delta(\kk+\kk'+\kk'') = - 2 {\mathrm i} \epsilon \Delta(\varOmega_{kk'k''}) \delta(\kk+\kk'+\kk'') \nonumber \\
&& \left[ L_{kk'k''}^{ss's''} q^{s''}_{k''}(\kk'') q^{s'}_{k'}(\kk') + L_{k'kk''}^{s'ss''} q^{s''}_{k''}(\kk'') q^{s}_{k}(\kk) 
+ L_{k''k'k}^{s''s's} q^{s}_{k}(\kk) q^{s'}_{k'}(\kk') \right] \, ;  \label{q4t3} 
\ea
then, eventually:
\ba
&&q^{ss's''}_{kk'k''}(\kk,\kk',\kk'') \delta(\kk+\kk'+\kk'') = - 2 {\mathrm i} \epsilon \Delta(\varOmega_{kk'k''}) \delta(\kk+\kk'+\kk'') \nonumber \quad \\
&& L_{kk'k''}^{ss's''}  \left[ q^{s''}_{k''}(\kk'') q^{s'}_{k'}(\kk') + ss' q^{s''}_{k''}(\kk'') q^{s}_{k}(\kk) + ss'' q^{s}_{k}(\kk) q^{s'}_{k'}(\kk') \right] \, . \label{q4t4}
\ea
The effective limit of long times (which introduces irreversibility) gives us (Riemann-Lebesgue lemma):
\be
\Delta(x) \to \pi \delta(x) + {\mathrm i} {\cal P} (1/x) \, , 
\ee
with ${\cal P}$ the principal value term. 
The so-called kinetic equation is obtained by injecting expression (\ref{q4t4}) into (\ref{2ndo}) and integrating over $\kk'$ (with the relation 
$q^{-s}_{-k}(-\kk)=q^{s}_{k}(\kk)$):
\ba
&&\frac{\partial q^{s}_{k}(\kk)}{\partial t} = 2 \epsilon^{2} \int \sum_{s_{p} s_{q}}  \vert L_{-kpq}^{-ss_{p}s_{q}} \vert^2
(\pi \delta(\varOmega_{-kpq}) + {\mathrm i} {\cal P} (1/\varOmega_{-kpq}))  {\mathrm e}^{i \Okpq t} \del \nonumber \\
&& s_{p}s_{q} \left[ s_{p}s_{q} q^{s_{q}}_{q}(\qq) q^{s_{p}}_{p}(\pp) - ss_{q} q^{s_{q}}_{q}(\qq) q^{s}_{k}(\kk) 
- ss_{p} q^{s}_{k}(\kk) q^{s_{p}}_{p}(\pp) \right]
{\mathrm d}\pp {\mathrm d}\qq \nonumber \\
&+& 2 \epsilon^{2} \int \sum_{s_{p} s_{q}}  \vert L_{kpq}^{ss_{p}s_{q}} \vert^2
(\pi \delta(\varOmega_{kpq}) + {\mathrm i} {\cal P} (1/\varOmega_{kpq})) {\mathrm e}^{i\varOmega_{kpq} t} \delta_{kpq} \nonumber \\
&& s_{p}s_{q} \left[ s_{p}s_{q} q^{s_{q}}_{q}(\qq) q^{s_{p}}_{p}(\pp) + ss_{q} q^{s_{q}}_{q}(\qq) q^{s}_{k}(\kk) 
+ ss_{p} q^{s}_{k}(\kk) q^{s_{p}}_{p}(\pp) \right]
{\mathrm d}\pp {\mathrm d}\qq \, . 
\ea
By changing the sign of the (mute) variables $\pp$, $\qq$, and the associated polarities, the principal value terms remove \citep{Benney1966}. 
Using the symmetries of the interaction coefficient, we finally arrive at the following expression after simplification:
\ba
\frac{\partial q^{s}_{k}(\kk)}{\partial t} &=& 
4 \pi \epsilon^{2} \int \sum_{s_{p} s_{q}} \vert L_{kpq}^{ss_{p}s_{q}} \vert^2 
\delta(s\omega_{k}+s_{p}\omega_{p}+s_{q}\omega_{q}) \delta(\kk+\pp+\qq) \nonumber \\
&& s_{p}s_{q}  \left[ s_{p}s_{q} q^{s_{q}}_{q}(\qq) q^{s_{p}}_{p}(\pp) + ss_{q} q^{s_{q}}_{q}(\qq) q^{s}_{k}(\kk) 
+ ss_{p} q^{s}_{k}(\kk) q^{s_{p}}_{p}(\pp) \right] {\mathrm d}\pp {\mathrm d}\qq \, , \label{KE1}
\ea
with:
\ba
&&\vert L_{kpq}^{ss_{p}s_{q}} \vert^{2}  \equiv \nonumber \\
&&\frac{\sqrt{\sigma}}{8} \left[s(\pp {\bm \cdot} \qq + pq) \left(\frac{pq}{k}\right)^{1/4} 
+ s_{p} (\kk {\bm \cdot} \qq + kq) \left(\frac{qk}{p}\right)^{1/4} + s_{q} (\kk {\bm \cdot} \pp + kp) \left(\frac{pk}{q}\right)^{1/4} \right]^{2} \, . 
\ea
Expression (\ref{KE1}) is the kinetic equation of capillary wave turbulence obtained for the first time by \citet{Zakharov1967}\footnote{To be totally convinced of the equivalence of the two expressions, it remains to develop the integrand according to the values of $s_p$ and $s_q$ by eliminating the particular case $s_p=s_q=s$ which has no solution at the resonance.}. 
Our writing in terms of directional polarity renders the expression more compact than in the original derivation.  
The presence of the small parameter $\epsilon \ll 1$ means that the amplitude of the quadratic nonlinearities is small and, therefore, the characteristic 
time that we consider to get a non-negligible nonlinear contribution is of the order of $1/\epsilon^2$. As we have seen, the derivation of this expression 
has been possible \ADD{because we have been able to manipulate and simplify the equations by using the symmetries of the interaction coefficient.}

It is from the inviscid invariants of the system that we can find the main properties of wave turbulence. The energy has, therefore, a privileged position 
since it is always conserved. Other (inviscid) invariants can be found as the kinetic helicity in incompressible hydrodynamics subjected to a rapid rotation 
(limit of weak Rossby numbers), a condition for being in the wave turbulence regime \citep{Galtier2003}. In the context of capillary wave turbulence, 
we will consider the only relevant invariant, the energy, for which the detailed conservation property will be proved. 
\ADD{Note that for four-wave interactions, like in gravity wave turbulence, the wave action is also conserved.}

%%%%%%%%%%%%%%%%%%
\section{Detailed energy conservation}\label{sec6}
%%%%%%%%%%%%%%%%%%

The kinetic equation (\ref{KE1}) describes the temporal evolution of capillary wave turbulence over asymptotically long times compared to the period 
of the waves. It is an equation involving three-wave interactions that give a non-zero contribution only when the following resonance condition is verified:
\begin{subequations}
\be
s\omega_{k}+s_{p}\omega_{p}+s_{q}\omega_{q} = 0 \, , \label{reson1}
\ee
\be
\kk+\pp+\qq = 0 \, . \label{reson2} 
\ee
\end{subequations}
For capillary waves, the resonance condition has solutions, but this is not always the case. For example, for gravity waves the dispersion 
relation $\omega_{k} \propto \sqrt{k}$ does not lead to a solution. In this case, it is necessary to consider the nonlinear contributions to the next order 
in development, i.e. the four-wave interactions; then, the problem becomes much more complicated \ADD{because one needs to go to higher 
order in the development which makes the writing even more complex and the calculation more lengthy.}

A remarkable property that verifies the kinetic equation is the detailed energy conservation. To demonstrate this result, we have to write the kinetic 
equation for the polarized energy:
\be
e^s(\kk) \equiv \omega_{k} q^{s}_{k}(\kk) \, .
\ee
We note in particular that: $e^s(\kk)=e^{-s}(-\kk)$. After some manipulations, we find:
\ba
\frac{\partial e^s(\kk)}{\partial t} &=& 
\frac{\pi \epsilon^{2}}{2\sigma} \int \sum_{s_{p} s_{q}} \vert \tilde L_{kpq}^{ss_{p}s_{q}} \vert^2 
\delta(s\omega_{k}+s_{p}\omega_{p}+s_{q}\omega_{q}) \delta(\kk+\pp+\qq) \quad \quad \nonumber \\
&& s \omega_{k} 
\left[ \frac{s\omega_{k}}{e^s(\kk)} + \frac{s_{p} \omega_{p}}{e^{s_{p}}(\pp)} + \frac{s_{q} \omega_{q}}{e^{s_{q}}(\qq)} \right] 
e^{s}(\kk) e^{s_{p}}(\pp) e^{s_{q}}(\qq) {\mathrm d}\pp {\mathrm d}\qq \, , \label{KE2} 
\ea
with:
\be
\vert \tilde L_{kpq}^{ss_{p}s_{q}} \vert^{2}  \equiv \left[\frac{\pp {\bm \cdot} \qq + pq}{sk\sqrt{pq}} 
+ \frac{\kk {\bm \cdot} \qq + kq}{s_{p} p\sqrt{kq}} + \frac{\kk {\bm \cdot} \pp + kp}{s_{q} q\sqrt{kp}} \right]^{2} \, .
\ee
Considering the integral over $\kk$ of the spectrum of the total energy, $e^+(\kk)+e^-(\kk)$, and then playing with the permutation of the wave vectors 
(in the first place, we divide the expression into three identical integrals), we can show that:
\ba
&&\frac{\partial \int \sum_s e^s(\kk) d\kk}{\partial t} = \frac{\pi \epsilon^{2}}{6\sigma} \int \sum_{s s_{p} s_{q}} \vert \tilde L_{kpq}^{ss_{p}s_{q}} \vert^2 
\delta(s\omega_{k}+s_{p}\omega_{p}+s_{q}\omega_{q}) \delta(\kk+\pp+\qq) \label{KE3} \nonumber \\
&& (s \omega_{k} + s_p \omega_{p} + s_q \omega_{q}) 
\left[ \frac{s\omega_{k}}{e^s(\kk)} + \frac{s_{p} \omega_{p}}{e^{s_{p}}(\pp)} + \frac{s_{q} \omega_{q}}{e^{s_{q}}(\qq)} \right] 
e^{s}(\kk) e^{s_{p}}(\pp) e^{s_{q}}(\qq) {\mathrm d}\kk {\mathrm d}\pp {\mathrm d}\qq \, \nonumber \\
&&= 0 \, . 
\ea
This means that the energy is conserved by triadic interaction: the redistribution of the energy is done within a triad satisfying the resonance condition 
(\ref{reson1})--(\ref{reson2}). It is a general property of wave turbulence that can be used to verify (in part) the accuracy of the kinetic equation obtained 
after a lengthy calculation.

%%%%%%%%%%%%%%%%%%
\section{Exact solutions and Zakharov transformation}\label{sec7}
%%%%%%%%%%%%%%%%%%

In this section, we look for the solutions of the kinetic equation (\ref{KE2}). A priori, this work is far from trivial and we can even wonder if it is possible 
to rigorously find exact  solutions from a nonlinear integro-differential equation. This is possible if we use a conformal transformation -- often called 
Zakharov transformation -- applied to the integral \citep{Zakharov1967}. It is curious to note that at the same time \citet{Kraichnan1967} proposed a 
similar transformation in the context of strong two-dimensional turbulence. 
\ADD{More recently, \cite{Balk2000} showed that exact solutions can also be derived without using a conformal transformation.}

\begin{figure}[t]
\center
\centerline{\includegraphics[width=0.4\linewidth]{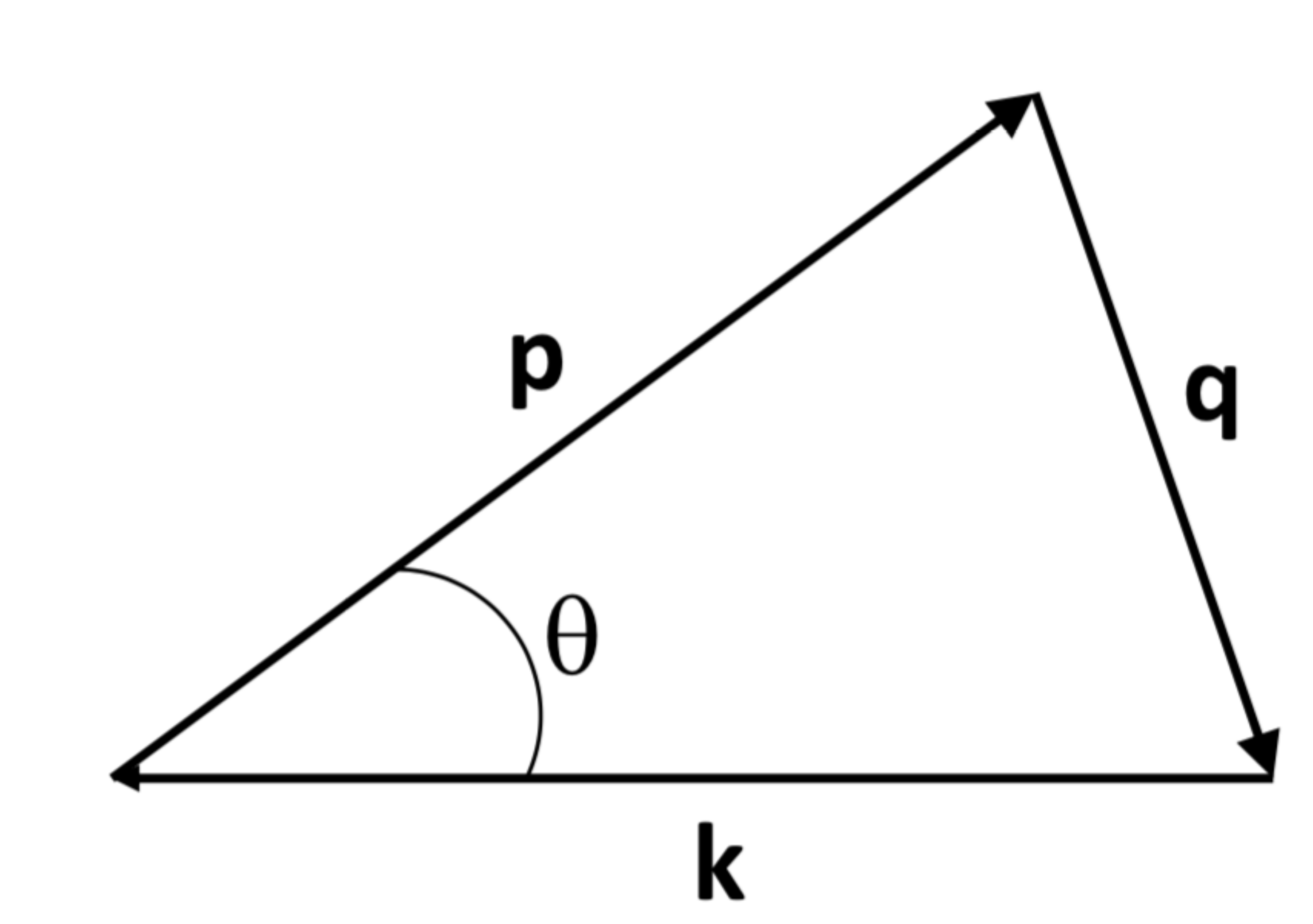}}
\caption{Triadic interaction.}
\label{Fig2}
\end{figure}
\begin{figure}[t]
\center
\centerline{\includegraphics[width=0.95\linewidth]{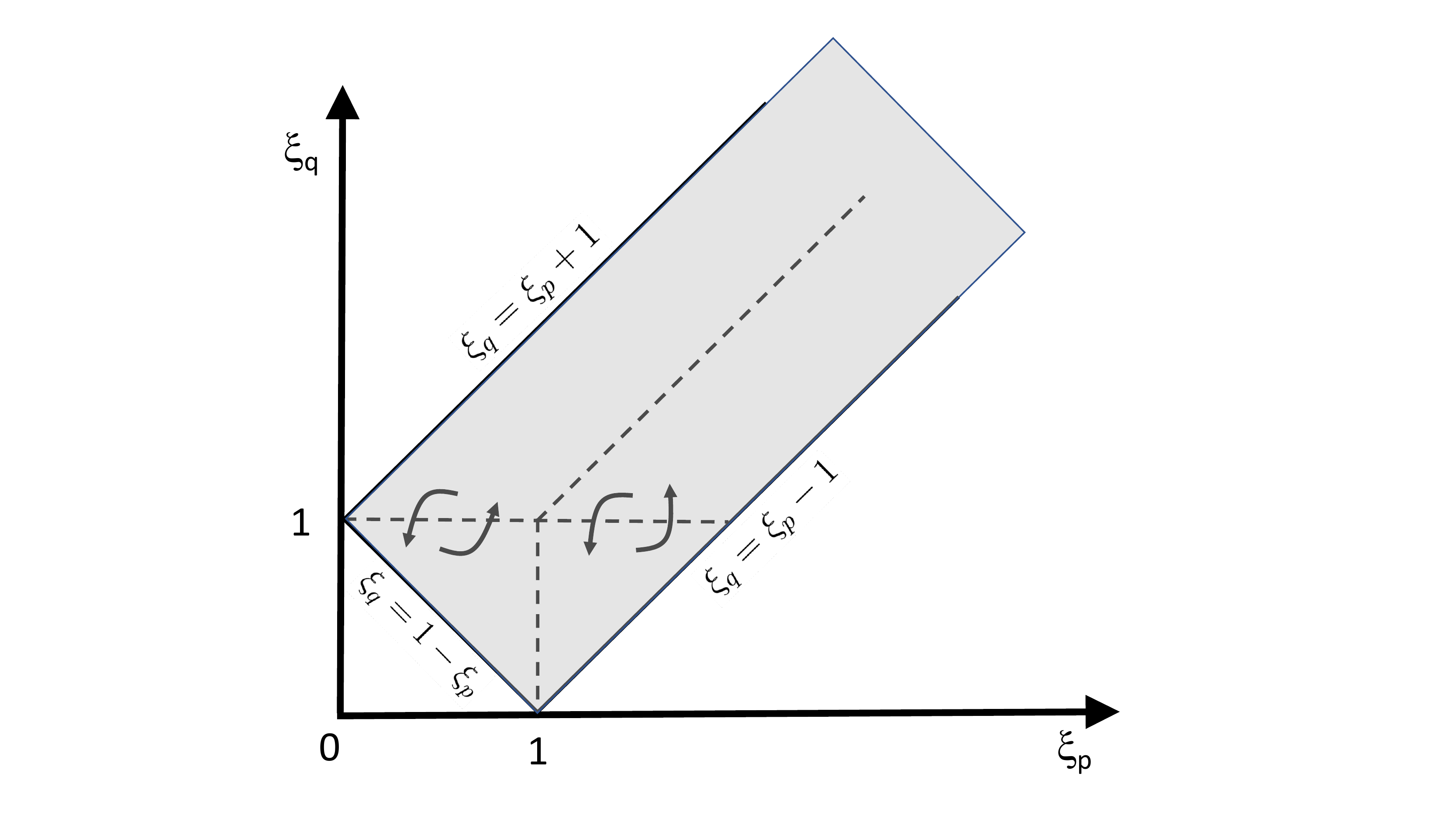}}
\caption{Conformal Zakharov transformation for triadic interactions. The gray band infinitely long corresponds to the solutions of the triangular relation 
$\kk+\pp+\qq={\bm 0}$. The transformation consists of the exchange of the four regions separated by the dashed lines: here, it is the transformation 
(\ref{TZ2}) which is visualized.}
\label{Fig3}
\end{figure}

We will make the simplifying hypothesis that capillary wave turbulence is statistically isotropic. This hypothesis is reasonable insofar as we have no 
source of anisotropy\footnote{\ADD{In nature or in numerical simulations several sources of anisotropy exist like the forcing or the inhomogeneities
due to the experimental setup. Here, this claim is made in the framework of the wave turbulence theory.}}. 
The situation can be different in other problems like in plasma physics where the strong uniform magnetic field that supports 
plasma waves induces also a strong anisotropy \citep[see e.g.][]{Meyrand2013}. Therefore, we introduce the isotropic spectrum:
\be
E(k) \equiv E_{k} = 2\pi k \sum_s e^s(\vert \kk \vert) \, .
\ee
We will rewrite the kinetic equation using the triangle relation (see figure \ref{Fig2}):
\be
q^{2} = k^2 + p^2 - 2 kp \cos \theta \, , 
\ee
from which we deduce (for fixed $k$ and $p$): $qdq = kp \sin\theta d \theta$. This relation will then be used to rewrite the kinetic equation.
After some manipulations we find:
\ba
&&\frac{\partial E_k}{\partial t} = \nonumber \\
&&\frac{\epsilon^{2}}{2 \sigma} \int_{\Delta} \sum_{s s_{p} s_{q}} s \omega_{k} \vert \tilde L_{kpq}^{s s_{p} s_{q}} \vert^2 
\delta(s \omega_{k}+s_{p}\omega_{p}+s_{q}\omega_{q}) 
\frac{sk\omega_{k} E_{p} E_{q} + s_{p} p \omega_{p} E_{k} E_{q} + s_{q} q \omega_{q} E_{k} E_{p}}
{\sqrt{4k^2p^2 -(k^2+p^2-q^2)^2}}  {\mathrm d}p {\mathrm d}q \, ,  \quad \quad
\ea
with: 
\be
\vert \tilde L_{kpq}^{ss_{p}s_{q}} \vert^{2}  = \left[\frac{k^2-p^2-q^2 + 2pq}{2sk\sqrt{pq}} 
+ \frac{p^2-k^2-q^2 + 2kq}{2s_{p} p\sqrt{kq}} + \frac{q^2-k^2-p^2 + 2kp}{2s_{q} q\sqrt{kp}} \right]^{2} \, .
\ee
Note that the new expression no longer uses wave vectors but only wave numbers; we also introduce the notation $\Delta$ for the integral to 
specify that the domain of integration is limited to triadic interactions, i.e. the gray band visible in figure \ref{Fig3}. 
An additional simplification is made by introducing the adimensionalized wave numbers $\xi_{p} \equiv p/k$ and $\xi_{q} \equiv q/k$; 
this gives the following expression:
\ba
\frac{\partial E_k}{\partial t} &=& \frac{\epsilon^{2} k^{5/2}}{2 \sqrt{\sigma}} 
\int_{\Delta} \sum_{s s_{p} s_{q}} s \vert \tilde L_{1\xi_{p}\xi_{q}}^{s s_{p} s_{q}} \vert^2 \delta(s+s_{p}\xi_{p}^{3/2}+s_{q}\xi_{q}^{3/2}) \nonumber \\
&&\frac{sE_{k\xi_{p}} E_{k\xi_{q}} + s_{p} \xi_{p}^{5/2} E_{k} E_{k\xi_{q}} + s_{q} \xi_{q}^{5/2} E_{k} E_{k\xi_{p}}}
{\sqrt{4\xi_{p}^2 -(1+\xi_{p}^2-\xi_{q}^2)^2}}  {\mathrm d}\xi_{p} {\mathrm d}\xi_{q} \, ,  
\ea
with: 
\be
\vert \tilde L_{1\xi_{p}\xi_{q}}^{ss_{p}s_{q}} \vert^{2}  = \left[\frac{1-\xi_p^2-\xi_q^2 + 2\xi_{p}\xi_{q}}{2s\sqrt{\xi_{p}\xi_{q}}} 
+ \frac{\xi_{p}^2-1-\xi_{q}^2 + 2\xi_{q}}{2s_{p} \xi_{p}\sqrt{\xi_{q}}} 
+ \frac{\xi_{q}^2-1-\xi_{p}^2 + 2\xi_{p}}{2s_{q} \xi_{q}\sqrt{\xi_{p}}} \right]^{2} \, .
\ee
We will apply the Zakharov transformation to this last expression by assuming a power law form for the energy spectrum, $E_{k} = C k^{x}$. 
In practice, we divide the integral into three equal parts and we apply on two of the three integrals a different transformation, keeping intact the 
third integral. These Zakharov transformations are:
\begin{subequations}
\be
\xi_{p} \to \frac{1}{\xi_{p}} \, , \quad \xi_{q} \to \frac{\xi_{q}}{\xi_{p}} \, , \quad \rm{(TZ1)} \label{TZ1}
\ee
and
\be
\xi_{p} \to \frac{\xi_{p}}{\xi_{q}} \, , \quad \xi_{q} \to \frac{1}{\xi_{q}} \, . \quad \rm{(TZ2)} \label{TZ2}
\ee
\end{subequations}
%%%%%%%%%%%%%%%%%%%%%%%%%%%%%%%%%%%
%%%%%%%%%%%%%%%%%%%%%%%%%%%%%%%%%%%
It corresponds to a conformal transformation of the region of integration in the space domain ($\xi_p$,$\xi_q$). 
In the case of triadic interactions, this region is an infinitely extended band as shown in figure \ref{Fig3}. It is straightforward to check that: 
\begin{subequations}
\be
\vert \tilde L_{1\xi_{p}\xi_{q}}^{s s_{p} s_{q}} \vert^2 \buildrel \rm TZ1 \over \longrightarrow \vert \tilde L_{1\xi_{p}\xi_{q}}^{s_{p} s s_{q}} \vert^2 \, , 
\ee
\be
\vert \tilde L_{1\xi_{p}\xi_{q}}^{s s_{p} s_{q}} \vert^2 \buildrel \rm TZ2 \over \longrightarrow \vert \tilde L_{1\xi_{p}\xi_{q}}^{s_{q} s_{p} s} \vert^2 \, , 
\ee
\be
\delta(s+s_{p}\xi_{p}^{3/2}+s_{p}\xi_{q}^{3/2}) \buildrel \rm TZ1 \over \longrightarrow \xi_{p}^{3/2} \delta(s_{p} + s \xi_{p}^{3/2}+s_{q}\xi_{q}^{3/2}) \, ,
\ee
\be
\delta(s+s_{p}\xi_{p}^{3/2}+s_{p}\xi_{q}^{3/2}) \buildrel \rm TZ2 \over \longrightarrow \xi_{q}^{3/2} \delta(s_{q} + s_{p} \xi_{p}^{3/2}+s\xi_{q}^{3/2}) \, , 
\ee
\be
1/\sqrt{4\xi_{p}^2 -(1+\xi_{p}^2-\xi_{q}^2)^2} \buildrel \rm TZ1 \over \longrightarrow \xi_{p}^2/\sqrt{4\xi_{p}^2 -(1+\xi_{p}^2-\xi_{q}^2)^2} \, , 
\ee
\be
1/\sqrt{4\xi_{p}^2 -(1+\xi_{p}^2-\xi_{q}^2)^2} \buildrel \rm TZ2 \over \longrightarrow \xi_{q}^2/\sqrt{4\xi_{p}^2 -(1+\xi_{p}^2-\xi_{q}^2)^2} \, , 
\ee
\be
{\mathrm d}\xi_{p} {\mathrm d}\xi_{q} \buildrel \rm TZ1 \over \longrightarrow \xi_{p}^{-3} {\mathrm d}\xi_{p} {\mathrm d}\xi_{q} \, , 
\ee
\be
{\mathrm d}\xi_{p} {\mathrm d}\xi_{q} \buildrel \rm TZ2 \over \longrightarrow \xi_{q}^{-3} {\mathrm d}\xi_{p} {\mathrm d}\xi_{q} \, .
\ee
\end{subequations}
The kinetic equation for the energy spectrum becomes:
\ba
\frac{\partial E_k}{\partial t} &=& \frac{\epsilon^{2} C^2 k^{2x+5/2}}{6 \sqrt{\sigma}} \int_{\Delta} \sum_{s s_{p} s_{q}}  
\frac{\vert \tilde L_{1\xi_{p}\xi_{q}}^{s s_{p} s_{q}} \vert^2}{\sqrt{4\xi_{p}^2 -(1+\xi_{p}^2-\xi_{q}^2)^2}} 
\delta(s+s_{p}\xi_{p}^{3/2}+s_{p}\xi_{q}^{3/2}) \nonumber \\
&&[\xi_{p}^x \xi_{q}^x ( s + s_{p} \xi_{p}^{-x+5/2} + s_{q} \xi_{q}^{-x+5/2}) s \nonumber \\
&&+\xi_{p}^x \xi_{q}^x (s + s_p \xi_{p}^{-x+5/2} + s_{q} \xi_{q}^{-x+5/2}) s_p \xi_{p}^{-2x-2}  \nonumber \\
&&+\xi_{p}^x \xi_{q}^x (s + s_p \xi_{p}^{-x+5/2} + s_{q} \xi_{q}^{-x+5/2}) s_q \xi_{q}^{-2x-2} 
]{\mathrm d}\xi_{p} {\mathrm d}\xi_{q} \, .  \label{KEso}
\ea
Note that to get the previous expression, we have exchanged $s$ and $s_p$ in the second term of the integral, and $s$ and $s_q$ in the third term 
of the integral. This manipulation is allowed because of the sum over the three indices $s$, $s_p$ and $s_q$.
\ADD{Note that the symmetric form of the kinetic equation is of great help in this part.}
After one last manipulation, we arrive at the following expression:
\ba
\frac{\partial E_k}{\partial t} &=& \frac{\epsilon^{2} C^2 k^{2x+5/2}}{6 \sqrt{\sigma}} \int_{\Delta} \sum_{s s_{p} s_{q}}  
\frac{\vert \tilde L_{1\xi_{p}\xi_{q}}^{s s_{p} s_{q}} \vert^2}{\sqrt{4\xi_{p}^2 -(1+\xi_{p}^2-\xi_{q}^2)^2}} 
\delta(s+s_{p}\xi_{p}^{3/2}+s_{p}\xi_{q}^{3/2}) \nonumber \\
&& \xi_{p}^x \xi_{q}^x [s +s_p \xi_{p}^{-x+5/2} + s_q \xi_{q}^{-x+5/2}][s + s_p \xi_{p}^{-2x-2} + s_q \xi_{q}^{-2x-2}] {\mathrm d}\xi_{p} {\mathrm d}\xi_{q} \, . 
\label{KEsolQ} 
\ea
The stationary solutions for which the term of the right-hand side vanishes correspond to:
\be
x = 1 \quad \rm{and} \quad x = -7/4 \, .
\ee
Indeed, for these two values the expression $s+s_{p}\xi_{p}^{3/2}+s_{q}\xi_{q}^{3/2}$ can emerge in the second line of (\ref{KEsolQ}), which 
vanishes exactly at the resonance (condition imposed by the Dirac). 

The solution $x=1$ corresponds to the thermodynamic equilibrium for which the energy flux is zero: in this case, each of the three terms in the right-hand side of 
(\ref{KEso}) vanishes and no transfer of energy is possible. The solution $x=-7/4$ is more interesting because the cancellation of the term in the right-hand side of 
(\ref{KEso}) is obtained by a subtle equilibrium of its three contributions: it is the finite flux solution called the Kolmogorov-Zakharov spectrum. In this case, 
a last calculation must be done to justify the relevance of this spectrum: it is a question of verifying the convergence of the integrals in the case of strongly 
non-local interactions. This corresponds to the regions close to the two right angles (see figure \ref{Fig3}) as well as the region infinitely distant from 
the origin. In the case of a divergence, the solution found is simply not relevant and only the numerical simulation of the wave turbulence equation makes 
it possible to estimate the shape of the spectrum. In capillary wave turbulence, the locality of the Kolmogorov-Zakharov spectrum has been checked by 
\citet{Zakharov1967}. Unlike the original derivation, here we have directly used the kinetic equation written in terms of wavenumbers, therefore the 
condition of locality must be checked carefully. Simple calculations leads to the condition:
\be
-2 < x < -3/2 \, ,
\ee
where the upper limit is fixed by the points at infinity in figure \ref{Fig3}. This result proves the relevance of the Kolmogorov-Zakharov spectrum. 
Note that the power law index of this solution is placed exactly at the middle of the convergence interval. This happens very often.

%%%%%%%%%%%%%%%%%%
\section{Nature of the spectral solutions}\label{sec8}
%%%%%%%%%%%%%%%%%%

The Zakharov transformation allowed us to obtain the exact stationary solutions in power law and to show two types of spectrum: the thermodynamic 
solution and the finite flux solution corresponding to a cascade of energy. However, the nature of this cascade remains unknown: is it a direct or an 
inverse cascade? The answer to this question needs to deepen our analysis.

We will use the relation that links the energy flux $\varepsilon_{k}$ to the energy spectrum:
\be
\frac{\partial E_k}{\partial t} = - \frac{\partial \varepsilon_k} {\partial k} = \frac{\epsilon^{2} C^2}{\sqrt{\sigma}} k^{2x+5/2} I(x)\, , 
\ee
where $I(x)$ is the integral deduced from expression (\ref{KEsolQ}). We obtain: 
\be
\varepsilon_k = - \frac{\epsilon^{2} C^2}{\sqrt{\sigma}} \frac{k^{2x+7/2} I(x)}{2x+7/2} \, . 
\ee
The direction of the cascade will be given by the sign of the energy flux in the particular case where $x=-7/4$. In this case, we see that the denominator
and the numerator cancel. With the l'Hospital's rule we find:
\ba
\lim_{x \to -7/4} \varepsilon_k &=& \varepsilon = - \frac{\epsilon^{2} C^2}{\sqrt{\sigma}} \lim_{x \to -7/4} \frac{I(x)}{2x+7/2} 
= - \frac{\epsilon^{2} C^2}{\sqrt{\sigma}} \lim_{y \to 3/2} \frac{I(y)}{3/2-y} \nonumber \\
&=& \frac{\epsilon^{2} C^2}{\sqrt{\sigma}} \lim_{y \to 3/2} \frac{\partial I(y)}{\partial y} \, , \label{ConsK}
\ea
with $y=-2x-2$ and:
\ba
\frac{\partial I(y)}{\partial y}\vert_{3/2} &=& \frac{1}{6} \int_{\Delta} \sum_{s s_{p} s_{q}} \nonumber
\frac{\vert \tilde L_{1\xi_{p}\xi_{q}}^{s s_{p} s_{q}} \vert^2}{\sqrt{4\xi_{p}^2 -(1+\xi_{p}^2-\xi_{q}^2)^2}} \delta(s+s_{p}\xi_{p}^{3/2}+s_{p}\xi_{q}^{3/2}) \\
&& \xi_{p}^x \xi_{q}^x [s +s_p \xi_{p}^{-x+5/2} + s_q \xi_{q}^{-x+5/2}] \frac{\partial [s + s_p \xi_{p}^{y} + s_q \xi_{q}^{y}] }{\partial y} 
{\mathrm d}\xi_{p} {\mathrm d}\xi_{q} \vert_{3/2}  
\,  \nonumber \\
&=& \frac{1}{6} \int_{\Delta} \sum_{s s_{p} s_{q}}  
\frac{\vert \tilde L_{1\xi_{p}\xi_{q}}^{s s_{p} s_{q}} \vert^2}{\sqrt{4\xi_{p}^2 -(1+\xi_{p}^2-\xi_{q}^2)^2}} \delta(s+s_{p}\xi_{p}^{3/2}+s_{q}\xi_{q}^{3/2})
\nonumber \\
&& \xi_{p}^{-7/4} \xi_{q}^{-7/4} [s +s_p \xi_{p}^{17/4} + s_q \xi_{q}^{17/4}] [s_p \xi_{p}^{3/2} \ln(\xi_{p}) + s_q \xi_{q}^{3/2} \ln(\xi_{q}) ] 
{\mathrm d}\xi_{p} {\mathrm d}\xi_{q} \, . \label{KEsolP}
\ea
The sign of the previous expression can be found numerically. A positive sign has been obtained \citep{Pushkarev2000} demonstrating that the energy 
cascade of capillary wave turbulence is direct. It is also possible to find the numerical value of the Kolmogorov constant $C_K$ whose expression is 
derived from $C$ in (\ref{ConsK}); we obtain finally:
\be
E_k = \frac{\sigma^{1/4}}{\epsilon} C_K \sqrt{\varepsilon} k^{-7/4} \quad \rm{with} \quad C_K =  \frac{1}{\sqrt{\partial I(y)/\partial y\vert_{3/2}}} \, . 
\ee
The value reported by \citet{Pushkarev2000} is $C_K \simeq 9.85$.

%%%%%%%%%%%%%%%%%%%%%%%%%%%%%%%%%%%%%%%%%%%%%%%%%%%%%%
\section{Experiments and simulations: a brief review}\label{sec9}
%%%%%%%%%%%%%%%%%%%%%%%%%%%%%%%%%%%%%%%%%%%%%%%%%%%%%%

Capillary waves have been studied for a long time, as demonstrated by the work of \citet{Longuet1963} or \citet{McGoldrick1970}: in these examples, 
the aim was to understand the role of resonant wave interactions as well as the mechanism of generation of capillary waves from gravity waves. 
On the other hand, the experimental study of capillary wave turbulence is more recent and is still the subject of many works today.
For example, \citet{Holt1996} have been able to produce a sea of capillary waves on the surface of a drop (of about $5$\,mm of diameter) in 
levitation and consisting mainly of water. The experiment shows the rapid formation (in less than a second) by a direct cascade of a spectrum in 
$f^{-3.58}$ (with $f$ the frequency) for the fluctuations of the surface elevation $\eta$. The difference with the theoretical prediction (\ref{KZsol4}) 
could be due to non-negligible viscoelastic effects.
In the article by \citet{Wright1996}, a new measurement technique based on the scattering of visible light on polystyrene spheres (with a diameter 
of the order of $\mu$m) is used to obtain the variations of the surface elevation of the water. From these data the authors were able to calculate the 
spectrum in wave numbers: a narrow power law with an exponent around $-4$ was measured. Note that subsequent measurements with the same 
technique (but with semi-skimmed milk) gave results consistent with the frequency prediction in $-17/6$ \citep{Henry2000}. 

Capillary wave turbulence was also studied from liquid hydrogen (maintained at $15-16$\,K). The interest of this type of study is that the liquid has 
a lower kinematic viscosity than water, thus making it possible to increase the size of the inertial zone \citep{Brazhnikov2001,Brazhnikov2002}. 
Note, however, that in this problem the inertial zone is also limited by the surface tension coefficient. The experiment shows that the theoretical  
frequency spectrum can be fairly well reproduced over more than one order of magnitude, despite significant noise 
\cite[see also][]{Kolmakov2004,Abdurakhimov2008}. 
Since the imagination of physicists is vast, the problem has also been tackled by using the fluorescence properties (located essentially on the surface 
of the liquid) of a solution added to water. With the help of a blue laser projected on the surface of the liquid, the authors \citep{Lommer2000} have 
accurately measured the power law and this time the Zakharov-Filonenko's solution has been very well reproduced over two decades of frequencies.

The low-gravity capillary wave turbulence regime ($\sim 0.05$\,g) has been achieved during parabolic flights (Airbus A320) of about $22$s \citep{Falcon2009}. 
The motivation for developing such an experience is to limit the effects of gravity waves and thus extend the inertial range of purely capillary waves. 
Two decades of power law in frequencies have been measured with an exponent close to the expected value. 
Note that gravity-capillary waves have been the subject of several experimental studies in which, in particular, the transition between large-scale gravity 
waves and small-scale capillary waves has been observed, as well as the possible (non-local) interactions between these two types of waves
\cite[see e.g.][]{Falcon2007,Deike2012,Berhanu2013,Aubourg2015,Aubourg2016,Berhanu2018,Hassaini2018,Cazaubiel2019}. 

Direct numerical simulation is a necessary complement to the experimental study because it allows a finer analysis of capillary wave turbulence: 
indeed the numerics gives access in principle to all fields and thus allows a precise data analysis. Few works, however, have been done in comparison 
to the many experimental studies. These simulations were first done by \citet{Pushkarev1996,Pushkarev2000}, then more recently by \citet{Deike2014} 
and \citet{Pan2014}.
Surface waves are, however, subject to numerical instabilities which make it difficult to obtain an extended inertial zone (less than one decade). 
Moreover, these simulations with an external force are particularly long to reach the stationary regime for which turbulence is fully developed. 
Finally, the Fourier space discretization induced by numerical simulation also has potential consequences on turbulence whose analytic properties 
are obtained in the context of a continuous medium \citep{Nazarenko11,Pan2017}. 
Paradoxically, the simulation of surface wave turbulence, by its two-dimensional nature, seems to be more difficult to make than that in three 
dimensions for which nonlinear interactions are more numerous and numerical instabilities better tamed.
Note that the works mentioned above complement the numerical simulations of \cite{Falkovich1995} realized from the wave turbulence equation in order 
to understand (for decaying turbulence) the non-stationary phase during which the spectrum propagates in an explosive way towards large wave numbers.

%%%%%%%%%%%%%%%%%%%%%%%%%%%%%%%%
\section{Conclusion}\label{sec10}
%%%%%%%%%%%%%%%%%%%%%%%%%%%%%%%%

Wave turbulence is a very active domain where the number of experiments has increased sharply over the past two decades. 
Nowadays, this turbulence regime is also accessible with direct numerical simulations and our understanding of wave turbulence, 
as well as its transition to strong turbulence, has been significantly improved. 
In the same time new areas of application have emerged, such as in cosmology with gravitational waves, and it is hard to imagine what will be the 
situation in a decade. In the case of capillary waves, it is surprising that since its first investigation in the sixties, this topic is always at the forefront 
of the current research, while it may be the simplest example of application of wave turbulence. For this reason it is important to better understand the 
theory of wave turbulence: this is precisely the main purpose of the present paper to contribute to this by giving for the first time the details of the 
derivation of the kinetic equations for capillary wave turbulence.

%%%%%%%%%%%%%%%%%%%%%%%%%%%%%%%%
\section*{Acknowledgments}
%%%%%%%%%%%%%%%%%%%%%%%%%%%%%%%%
I would like to thank the organizers of the summer school of Carg\`ese, {\it Waves, Instabilities and Turbulence in Geophysical and Astrophysical Flows 
-- 2019}, for their invitation that allowed me to address the problem of capillary wave turbulence. 

%\bibliography{MyBiblio}

\end{document}